\documentclass[prl,twocolumn,amsmath,amssymb]{revtex4}

\usepackage{amsmath}
\usepackage{amssymb}
\usepackage{graphicx}

\bibliography{ref}

\usepackage{float}

\usepackage[T1]{fontenc}
\usepackage{indentfirst}
\begin{document}
\title{Distinct Magnetic Phases in Structurally Uniform SrCoO$_{3-y}$ }
\author{Z. H. Zhu,$^1$ F. J. Rueckert,$^1$ J. I. Budnick,$^{1,2}$ W. A. Hines,$^1$  Ch. Niedermayer,$^3$ L. Keller,$^3$ H. Luetkens,$^4$ B. Dabrowski,$^5$ S. Kolesnik,$^5$ B. O. Wells$^{1, 2}$ \\ $^1$Department of Physics, University of Connecticut, Storrs, Connecticut 06269-3046, USA\\ $^2$Institute of Material Science, University of Connecticut, Storrs, Connecticut 06269-3136, USA\\$^3$Laboratory for Neutron Scattering and Imaging, Paul Scherrer
Institute, CH-5232 Villigen, Switzerland\\$^4$ Laboratory for Muon Spin Spectroscopy, Paul Scherrer
Institute, CH-5232 Villigen, Switzerland\\
$^5$Physics Department, Northern Illinois University, Dekalb, Illinois 60115, USA}
\date{\today}

\begin{abstract} Two magnetic phase transitions have been noted for SrCoO$_{3-y}$ for near-stoichiometric oxygen concentrations (small y). Using muon spin rotation and neutron scattering experiments, we have established that the two transitions represent separate, spatially distinct magnetic phases that coexist in a two-phase equilibrium mixture. The two phases most likely represent areas of the sample with different effective valence charge density. Further, the phases exist over regions with a length scale intermediate between nanoscale charge inhomogeneity and systems such as manganites or super-oxygenated cuprates with large length scale phase separation.
\end{abstract}
\maketitle

There is a growing body of evidence that the ground state of strongly correlated, transition metal oxides doped away from half-filling may be characterized by electronic phase separation, i.e. that different electronic ground states coexist at low temperature in a given sample [1]. This is an aspect of the competition between differing phases with small differences in free energy that are prevalent in these compounds and plays a role in the variety of properties observed. The best known examples of such phase separation are the manganite compounds with colossal magnetoresistance (CMR). The CMR itself is a result of competition between an insulating antiferromagnetic phase and a conducting ferromagnetic phase [2]. In cuprates, electronic inhomogeneity at short length scales is well documented [3-9] but actual large length scale phase separation occurs for the special case of superoxygenated La$_{2-x}$Sr$_x$CuO$_{4+y}$. Phase separation in this case involves distinct magnetic and superconducting regions with different effective hole densities [10-15].

The phase diagram of cobaltite La$_{1-x}$Sr$_x$CoO$_{3-y}$ exhibits many similar properties to the cuprates. Hole doping LaCoO$_3$ through substitution of Sr$^{2+}$ for La$^{3+}$ leads to a  spin glass and short length scale charge variation [16-23]. Controlling the charge doping of SrCoO$_{2.5}$ - SrCoO$_3$ with mobile oxygen defects leads to the coexistence of multiple magnetic phases.  For example, SrCoO$_{2.88}$ is ferromagnetic with $T_C$ = 220 K and SrCoO$_3$ is ferromagnetic with $T_C$ = 280 K. For intermediate oxygen concentrations a double transition appears with characteristic temperatures of 220 K and 280 K, despite the fact that the crystal structure is intermediate between SrCoO$_{2.88}$ and SrCoO$_3$ and not a sum of these end point structures[24]. While this may reflect a magnetic phase separation, no local probes have examined the spatial variation of the magnetism. Here we report just such a study. Muon spin rotation ($\mu^+$SR) measurements reveal mixed phase samples that consist of spatially separated magnetic phases closely related to those end point compounds. The details of the muon spectra imply the separate regions have a length scale intermediate between samples with nanoscale electronic inhomogeneity and others with large length scale phase separation. The results reported here have important implications for understanding electronic phase separation itself. One is that we find the phenomenon in a material without exotic properties such as superconductivity or colossal magneto-resistance. In fact, the separate phases are similar ferromagnetic states. This implies that electronic phase separation may be an inherent feature of doped Mott insulators generally rather than being associated with any particular, exotic state. A second issue is that of length scale, which has not previously been a focus of the literature on electronic phase separation. The discovery of a material with an intermediate length scale establishes a link between the apparently different phenomena of two phase mixture-like behavior seen in a few extraordinary systems and the more common observation of nanoscale variations in the local charge density.

Our starting parent cobaltite SrCoO$_{2.88}$ was synthesized using a conventional solid reaction process along with high-pressure oxygen annealing [25]. Electrochemical oxidation was used to alter the oxygen concentration between the values of SrCoO$_{2.88}$ and SrCoO$_3$. Measurements of dc magnetization were performed on a Quantum Design MPMS SQUID to identify the magnetic phases. $\mu$$^+$SR measurements were carried out using the General Purpose Surface-Muon (GPS) Instrument at the Paul Scherrer Insitute. Neutron measurements were carried out using the cold neutron powder diffractometer DMC at the Swiss Spallation Neutron Source SINQ, Paul Scherrer Insitute.

\begin{figure}[htb]
\centering
\includegraphics[scale=0.45]{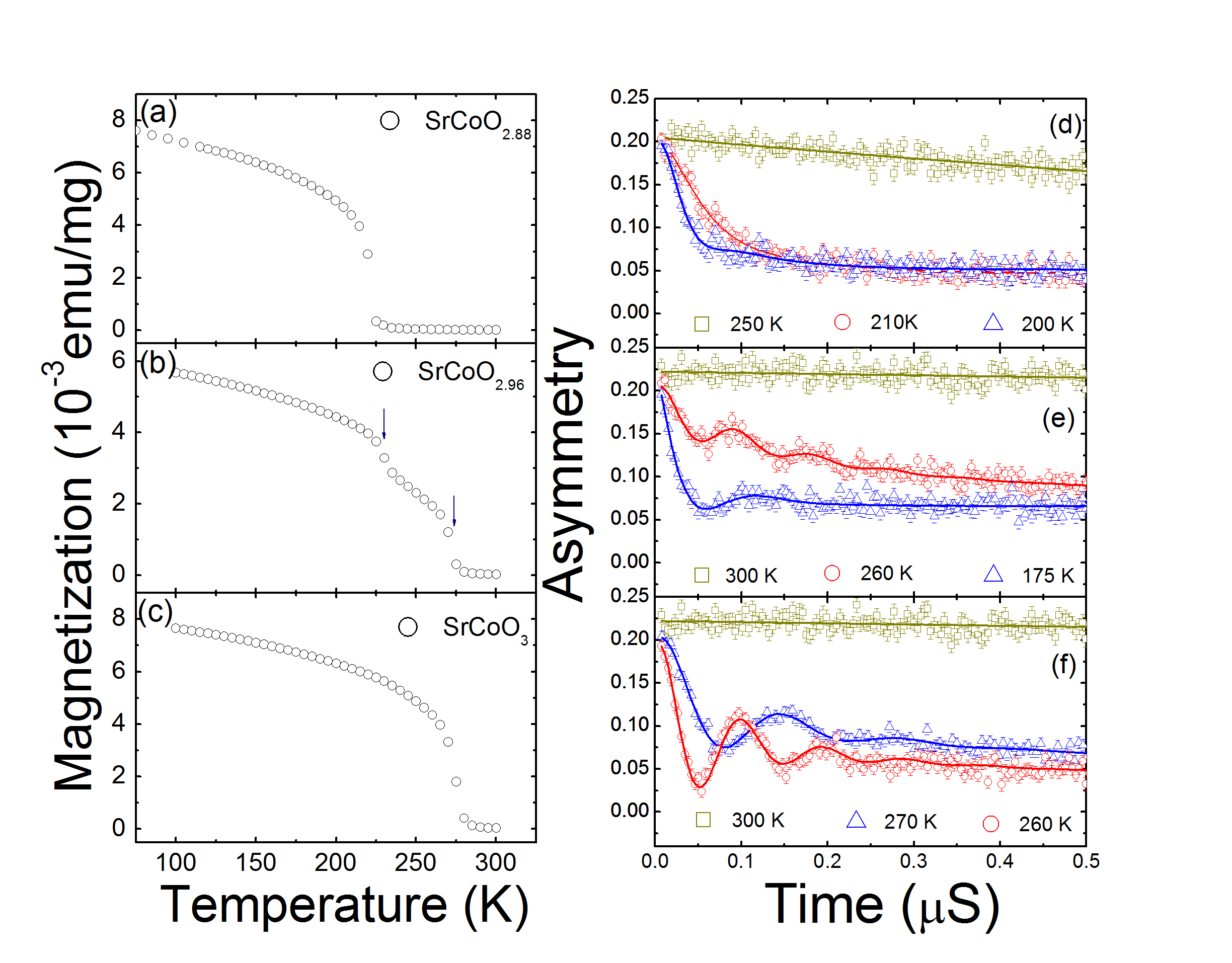}
\caption{\label{fig:epsart}(a), (b) and (c) are field-cooled(FC) magnetization versus temperature for a magnetic field H = 50 Oe for SrCoO$_{2.88}$, SrCoO$_{2.96}$, and SrCoO$_3$, respectively. (d), (e), and (f) are  zero field (ZF-)$\mu$$^+$SR time spectra for SrCoO$_{2.88}$, SrCoO$_{2.96}$, and SrCoO$_3$, respectively, at various temperatures. Solid lines represent fits using Eq.(1) for SrCoO$_3$ and SrCoO$_{2.88}$, respectively. Solid line in Fig. 1(e) represents fit using Eq. (2). }
\vspace{-0.0cm}
\end{figure}

In Fig. 1 we show the muon data that is the essential result of this work. The left side panels, Figs. 1(a), (b), and (c) are magnetization versus temperature for SrCoO$_{2.88}$, SrCoO$_{2.96}$, and SrCoO$_3$, respectively. The two transitions highlighted by arrows in Fig. 1(b) for partially oxidized SrCoO$_{2.96}$ indicate there are two magnetic phase transitions matching those found in parent SrCoO$_{2.88}$ and stoichiometric SrCoO$_3$ with $T_C = 220 K$ and $280 K$, respectively. The panels on the right side, Figs. 1(d), (e) , and (f) are representative $\mu$$^+$SR time spectra in zero applied field for the same samples measured in panels (a), (b), and (c). In the ferromagnetic state below $T_C$ internal fields at the muon site develop and therefore a spontaneous precession of the muon spin is observed. In a powder sample $\frac{2}{3}$ ($\frac{1}{3}$) of the field components are perpendicular (parallel) to the initial muon spin polarization resulting in the following time evolution of the polarization[15]: \begin{equation}
P(t)=\frac{A(t)}{A_0}= \frac{2}{3}cos(2\pi \nu_\mu t)exp(-\lambda_{osc} t)+ \frac{1}{3}exp(-\lambda_{tail}t), 
\end {equation}
where $ P(t)$  is the muon spin polarization function. $A(t)$ and $A_0$ are asymmetry and total initial asymmetry, respectively. The $\nu $$_\mu $ is the muon-spin precession frequency due to the internal field at the muon sites. The damping rate $\lambda$$_{osc}$ is dominated by the static distribution of the local field. The $\lambda$$_{tail}$ reflects the spin lattice relaxation and $\lambda$$_{tail}$ = 0 in the static case. In Fig. 1(f) clear oscillations for SrCoO$_3$ appear below 280 K with an amplitude reflecting the powder average, i.e. the entire sample is magnetic. As shown in Fig 1(d), no clear oscillations are observed in the asymmetry spectra for SrCoO$_{2.88}$ though the substantial relaxation that sets in below 220K implies that the entire sample becomes magnetic, albeit with a greater variation in local fields than in the SrCoO$_3$ sample.

Data for the mixed phase sample between $220 K$ and $280 K$, however, cannot be well fit using the simple formula above, indicating a more complex magnetic structure. For example,  at $260 K$ as shown in  Fig. 1(e), the ZF time spectrum  essentially consists of four components, oscillating term, very rapid damping, slow relaxation, and $\frac{1}{3}$ tail term. Therefore, the formula with four components is constructed as following:
\begin{equation}
\begin{split}
P(t)=  \frac{2}{3}[A_{osc}exp(-\lambda_{osc}t)cos(2\pi \nu_\mu t)+\\A_{fast}exp(-\lambda _{fast}t)+A_{slow}exp(-\lambda _{slow}t)]\\+ \frac{1}{3} exp(-\lambda_{tail}t), 
\end{split}
\end{equation}
where A$_{osc}$+A$_{fast}$+A$_{slow}$ = 1. A$_{osc}$, $\lambda_{osc}$, and $\nu_\mu$ are the fraction, relaxation rate, and oscillation frequency of the oscillating component, respectively. A$_{fast}$  and $\lambda _{fast}$ are the fraction and relaxation rate for the  rapid damping component, while A$_{slow}$  and $\lambda _{slow}$ represent the fraction and depolarization rate from the slow damping contribution. 
 \begin{figure}[htb]
\begin{center}
\includegraphics [scale =.290]{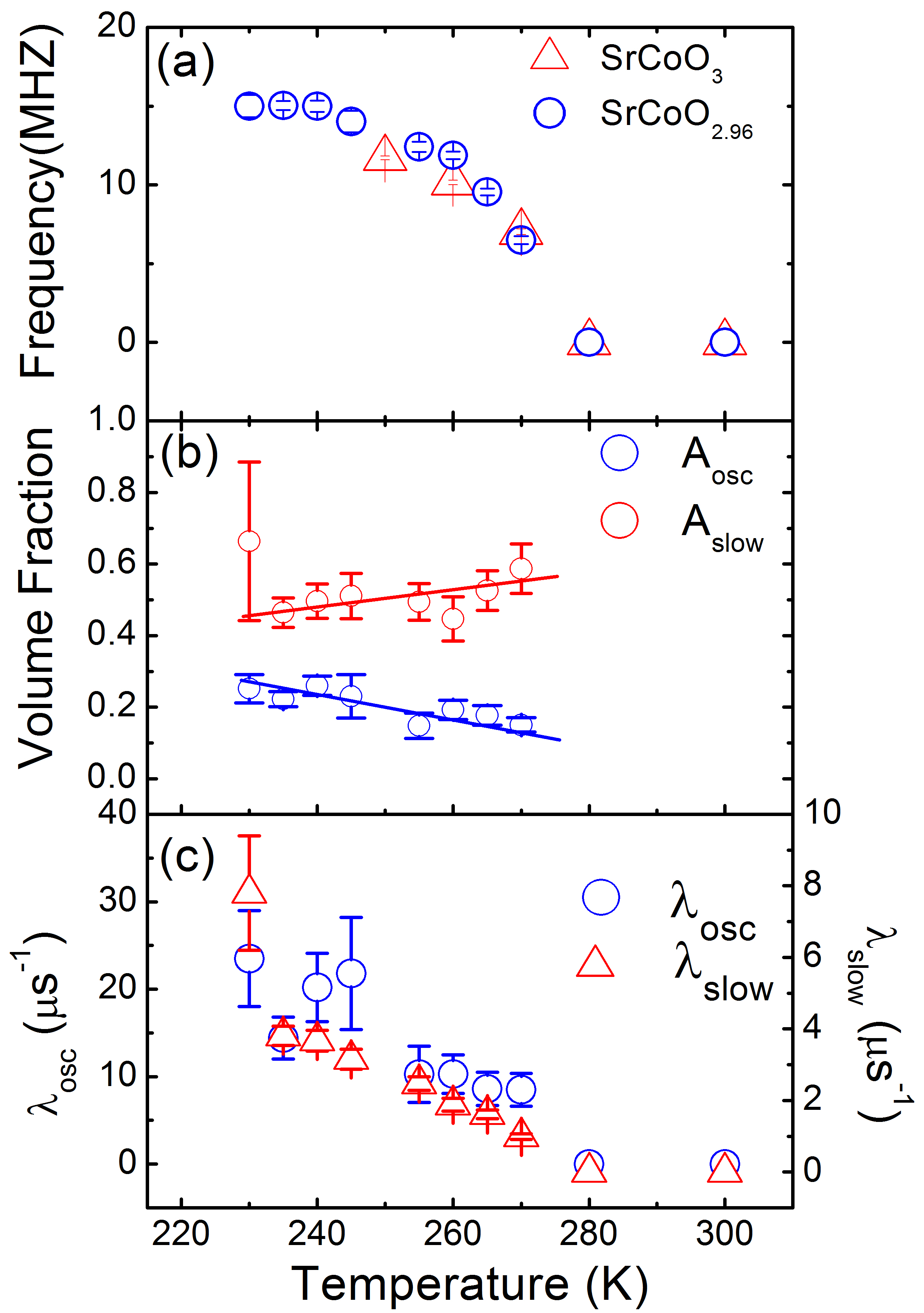}
\end {center}
\caption{\label{fig:epsart}(a)Temperature dependence of muon processing frequency for SrCoO$_{2.96}$ and SrCoO$_{3}$. (b) Temperature dependence of the volume fractions from the oscillation phase(A$_{osc}$) and the slow damping phase(A$_{slow}$) for SrCoO$_{2.96}$.(Solid lines are guides to the eye) (c) Temperature dependence of the relaxation rates of  oscillation phase ($\lambda_{osc}$) and slow damping phase $\lambda_{slow}$).  }
\vspace{0.0cm}
\end{figure}

 \begin{figure}[htp]\includegraphics[scale=0.45]{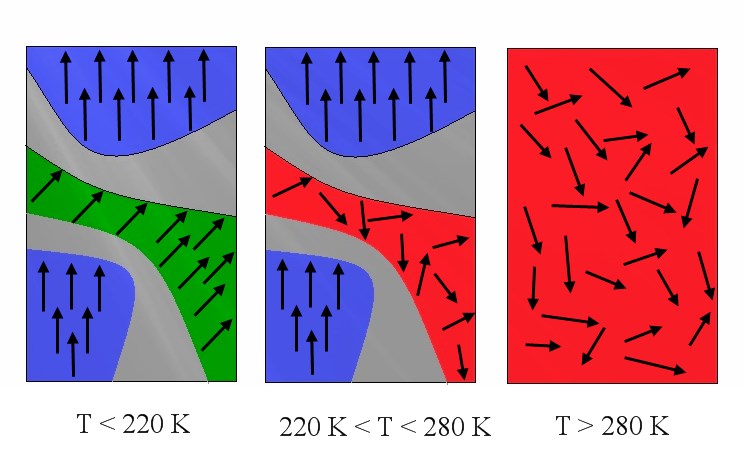}
\centering
\caption{\label{fig:epsart}A schematic plot of the micro-structure of the magnetic order in the SrCoO$_{2.96}$ for different temperature ranges. Above the $280 K$, the whole sample is not ordered yet;in between $220 K$ and $280 K$ the sample consists of spatially separated ordered and disordered regions, as well as the boundary symbolized by the gray area; below $220 K$, both two phases are magnetically ordered. } 
\vspace{0.0cm}
\end{figure}
Key parameters from fits to the data set of the mixed phase sample in between $220 K$ and $280 K$ are shown in Fig. 2. Fig. 2(a) plots the frequency as a function of temperature, with the frequency from the SrCoO$_3$ sample included as a reference. The two frequencies are identical within the error bars in this temperature range, both growing in a manner characteristic of an order parameter with a critical temperature of $280 K$. As shown in Fig. 2(b), the volume fraction A$_{osc}$ of this oscillating phase increase from about 20\% to 30\% of the sample when decreasing temperature from $280 K$ to $220 K$. The volume fraction of the slow damped contribution, however, decreases accordingly with decreasing temperature. The combination A$_{osc}$+A$_{slow}$ is almost constant and occupies about 72\%. Therefore, the contribution from the rapid damping phase is about 28\%, a significant amount. Additionally, in Fig. 2(c) the depolarization rates of the oscillation ( $\lambda_{osc}$ ) and the slow damping phase($\lambda_{slow}$) behave like  order parameters as well, and give the same $T_C$ at $280 K$. The static magnetic inhomogeneity in both regions scales with the size of the ordered moment that drives the oscillations; thus must be driven by the same fields.

The zero-field muon results allows us to construct a model for the spatial magnetic structure in the intermediate compound.The oscillating term represents ordered magnetic regions with uniform local moments. The term with moderate relaxation arises from paramagnetic regions of the sample, with some increase in relaxation rate due to stray fields from the ordered region. This leaves the highly damped signal to be from boundary regions, where the strong damping indicates a static, inhomogeneous magnetic state that forms a boundary between magnetic and paramagnetic regions. A schematic plot of the micro structure of the magnetic order is proposed in Fig. 3, showing proposed structures for temperatures above both transitions, between the upper and lower transition, and at lower temperatures where both phases are ordered. Further experiments, described below, provide support for this model.  

Transverse field muon (TF-$\mu^+$SR) measurements are helpful in confirming the origin of the slowly relaxing signal. In TF-$\mu^+$SR muons stopping in paramagnetic regions will precess with a frequency determined by the external magnetic field(50 Oe). Muons stopping in magnetically ordered regions will see a field determined by the sum of the local internal field and the applied field, which typically varies strongly due to the local field orientation and thus gives a strongly damped signal. We fit the data with a combination of an exponentially dampened cosine oscillation due to the external field for the nonmagnetic region and an exponentially relaxing non-oscillatory contribution from the randomly distributed internal fields in the ordered states as in the following equation: \begin{figure}[htp]\includegraphics[scale=0.45]{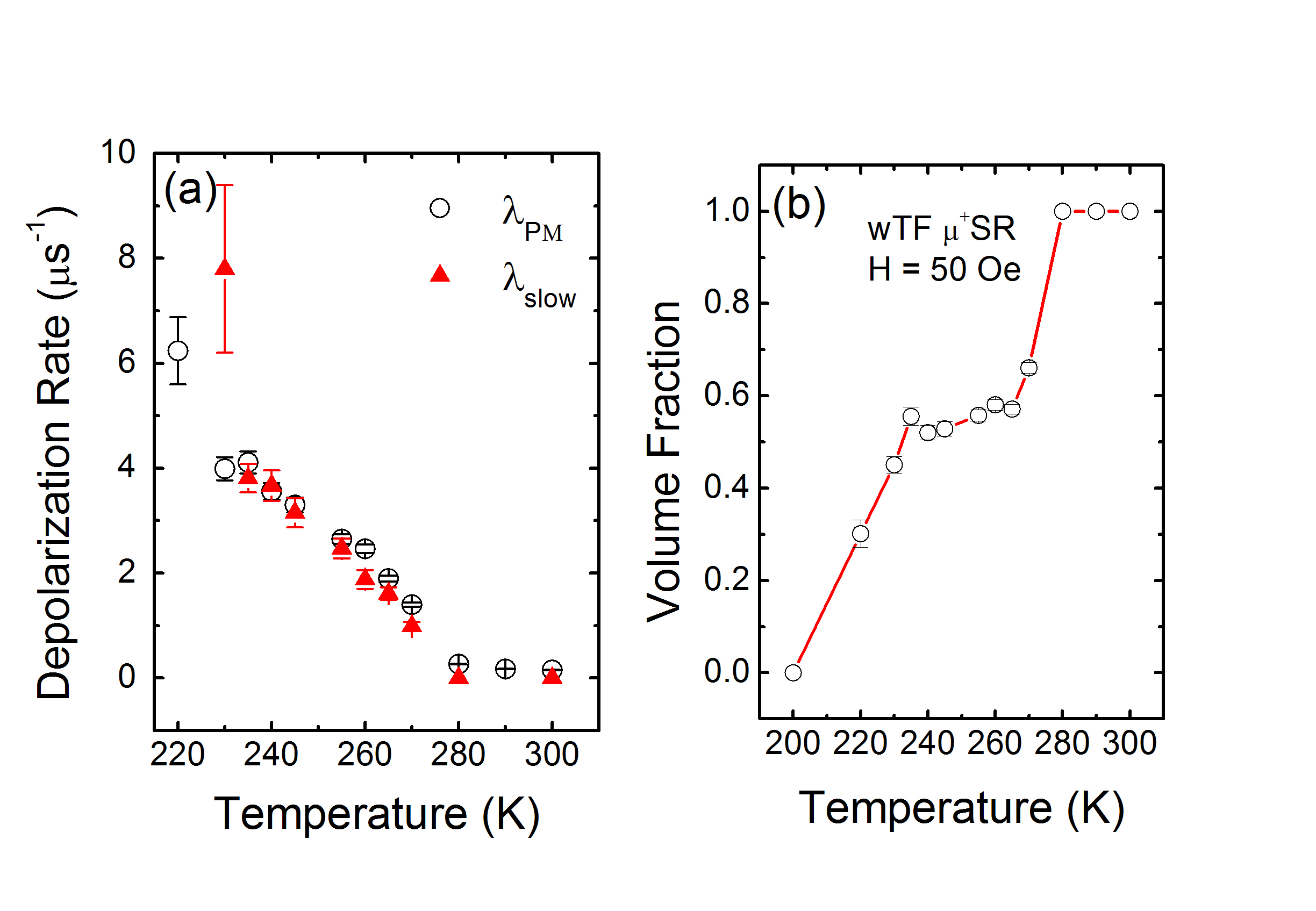}
\caption{\label{fig:epsart}(a)Temperture dependence of the relaxation rates of the slow damping phase(red solid triangle) measured from the ZF-$\mu$$^+$SR and the paramagnetic phase(open circle) measured from the Transverse Field(TF)-$\mu$$^+$SR. (b)Temperature dependence of the volume fraction of the nonmagnetic regions obtained by using the Transverse Field(TF)-$\mu$$^+$SR. The solid line is a guide to the eye.} 
\vspace{0.0cm}
\end{figure}
\begin{equation}
P(t)= A_{PM}exp(-\lambda_{PM}t)cos(\omega _\mu t+\phi)+A_Mexp(-\lambda _M t),
\end{equation}
where $A_{PM}+A_M=1$. $\omega$$_\mu$ is the muon Larmor frequency from the applied field, $\phi$ is the initial phase of the muon precession, and $A_{(PM or M)}$ and $\lambda_{(PM or M)}$ are asymmetries and relaxation rates of the nonmagnetic  and magnetic phases. In Fig. 4(a) we show the resulting relaxation rate from the paramagnetic region $\lambda_{PM}$ as a function of temperature. This is plotted along with the parameter labeled as the slow relaxation rate from the ZF-$\mu^+$SR measurement which is also shown in Fig. 2(c). The fact that these rates match is evidence that the signal labeled as ``slow-relaxation'' in the ZF-$\mu^+$SR is in fact most likely from paramagnetic regions \textendash with the damping rate given by the range of stray fields in both cases. Further, the temperature dependence of the phase fraction for this paramagnetic region is given in Fig. 4(b). In the intermediate temperature regime this comprises a little over 50\% of the sample, roughly matching the phase fraction of the slow-relaxing signal in the ZF data. Below the lower transition, the volume fraction of the paramagnetic regions shrinks to zero at 200 K, implying that the whole sample is magnetically ordered though inhomogeneous. Taken together there is strong evidence that the both the slowly relaxing term in the ZF asymmetry and the oscillating term in the TF spectra both represent the same, paramagnetic fraction of the sample.

\begin{table}
\caption{\label{tab:table1}Refined magnetic moment of Co ions for SrCoO$_3$, SrCoO$_{2.96}$, and SrCoO$_{2.88}$, respectivey, obtained by using neutron powder diffraction. }
\begin{ruledtabular}
\begin{tabular}{lccr}
Sample&250 K\ &150 K\ &1.5 K\\
\hline
SrCoO$_3$ & 1.0(1)$\mu$$_B$& 1.6(1)$\mu$$_B$&1.7(1)$\mu$$_B$\\
SrCoO$_{2.96}$ & 0.65(6)$\mu$$_B$ & 1.6(1)$\mu$$_B$&1.7(1)$\mu$$_B$\\
SrCoO$_{2.88}$ & Not Ordered& 1.3(1)$\mu$$_B$&1.7(1)$\mu$$_B$\\
\end{tabular}
\end{ruledtabular}
\end{table} 

We have also performed neutron powder diffraction measurements on similarly prepared and oxygenated samples. Table I summarizes the ordered moments found from fitting the spectra and assuming the full volume contributes to the peak intensities. In each case the diffraction patterns at low temperature are consistent with a ferromagnetic state, though there are some small unidentified peaks. In the mixed phase compound, below the second transition at $220 K$, there are no new peaks or broadening of peaks that would indicate a magnetic state with a different periodicity or riding on a different lattice than that which sets in at $280 K$. However, the magnetic moment that appears in the mixed phase sample below $280 K$ is quite a bit smaller than that at lower temperatures or in SrCoO$_3$ at the same temperature. This is consistent with only a fraction of the sample becoming magnetic below $280 K$. From the muon frequencies, we can assume that the moment per Co ion in the part of the mixed phase sample that is magnetic between $220 K$ and $280 K$ matches that from SrCoO$_3$. With that assumption, the volume fraction of magnetic ordered region can be estimated assuming $I \propto Vm^2$, in which $I$ is the magnetic neutron intensity; $V$ is the volume of magnetic ordered region and $m$ is the magnetic moment. A magnetic moment from the full volume of 0.65(6) $\mu_B$ in the mixed phase sample at $250 K$ would also be consistent with a moment of 1.0 $\mu_B$ throughout 40\% to 50\% of the volume. While neutrons do not allow for directly measuring the spatial variation of magnetism within a sample, the results are consistent with the picture of spatial phase separation derived from the muon data.

The separate magnetic regions seem most likely determined by the effective charge density. The regions with $T_C=280K$ appear to have the same magnetic state as SrCoO$_3$, and thus likely the same effective hole density; consisting of all Co$^{4+}$ ions. Similarly, the regions with T$_C= 220K$ appears to have the same magnetic state as SrCoO$_{2.88}$, presumably with the same effective charge density or an average Co valence of $\emph{+3.875}$. Normally we would expect such differing charge densities to involve a large cost in Coulomb energy, however since the key aspect of samples exhibiting separate magnetic states appears to be mobility of the dopant ions, we speculate that the counter ions have also reordered leaving the actual charge of the separate regions neutral.

Probably the most interesting result of this study compared to other examples of electronic phase separation is that the regions appear to exist at an intermediate length scale. The most common case is nano-scale charge inhomogeneity or separation. The different magnetic phases in our case must be larger than those as they support distinct magnetic transitions and two-phase behavior. In fact, $T_C$ for each region is not significantly reduced from the end points, indicating that there is no finite size related reduction of the ordering temperature. This means the separate magnetic regions must be much larger than the characteristic spin-spin correlation length, which was determined to be about nine lattice constants in a study of La$_{0.5}$Sr$_{0.5}$CoO$_3$ films [26]. In addition, the size of the separate regions in these cobaltites appears smaller than in some other systems known to show electronic phase separation. For example, superoxygenated La$_{2-x}$Sr$_x$CuO$_{4+y}$ separates into superconducting and stripe-like magnetic domains with regions that appear to be quite large; the muon asymmetry histograms can be fit by a simple sum of the magnetic and non-magnetic regions without either a significant contribution from a border region or a large change in damping of the non-magnetic region [9,11]. In the cobaltite case, both adjustments are necessary, indicating that the length scale of the electronic phases is on the order of the local dipole field range , a value considered to be several tens of nanometers [27]. The overall picture that emerges is that the phenomenon of electronic phase separation is strongly governed by length scales: frozen dopant ions lead to short length that appear in most experiments as local charge inhomogeneity within in a single electronic phase whereas highly mobile dopant ions allow for large length scale phase separation with two well-defined charge states. The details of material parameters allow for intermediate cases. SrCoO$_{2.5}$ is an insulating antiferromagnet whereas for oxygen concentrations closer to 3 the material becomes conducting with no reported deviations at the Curie temperature.

In conclusion, we report the direct evidence of phase separation in SrCoO$_{3-y}$ with 0 < y < 0.12 by means of muon spin rotation and powder neutron diffraction measurements. This report adds to the few known cases of doped Mott insulators exhibiting equilibrium separation into two stable charge states. It also demonstrates the importance of length scale in discussing such phenomena as this system has large enough separated regions to exhibit clear magnetic properties but are small enough that interaction regions substantially effect the spectra of local probes.

This work is based on experiments performed at the Swiss Muon Source (S$\mu$S) and the Swiss spallation neutron source SINQ, Paul Scherrer Institute, Villigen, Switzerland.
Sample synthesis at the University of Connecticut was supported by NSF
through grant DMR-0907197, muon and neutron data collection and analysis by
DOE-BES Contract No. DE-FG02-00ER45801.  Research at Northern Illinois University was supported by the Institute for Nanoscience, Engineering, and Technology - InSET.

\end{document}